# Strong electrostatic control of excitonic features in MoS$_2$ by a free-standing ultrahigh-κ ferroelectric perovskite

*Thomas Pucher[1*], Sergio Puebla[1], Victor Zamora[2], Estrella Sánchez Viso[1], Victor Rouco[2], Carlos Leon[2,3], Mar Garcia-Hernandez[1,3], Jacobo Santamaria[2,3], Carmen Munuera[1,3*] and Andres Castellanos-Gomez[1,3*]*

[1] *2D Foundry research group. Instituto de Ciencia de Materiales de Madrid (ICMM-CSIC), Madrid, 28049, Spain.*

[2] *GFMC, Department Fisica de Materiales, Facultad de Fisica, Universidad Complutense 28040 Madrid, Spain*

[3] *Unidad Asociada UCM/CSIC, "Laboratorio de Heteroestructuras con aplicación en spintrónica"*

thomas.pucher@csic.es

cmunuera@icmm.csic.es

andres.castellanos@csic.es

## ABSTRACT

We present the electrostatic control of photoluminescence of monolayer MoS$_2$ at room temperature via integration of free-standing BaTiO$_3$ (BTO), a ferroelectric perovskite oxide, layers. We show that the use of BTO leads to highly tunable exciton emission of MoS$_2$ in a minimal range of gate voltages, effectively controlling the neutral excitons to charged excitons (trions) conversion. Due to BTO's ferroelectric polarization-induced doping we observe large peak emission shifts as well as a large and tunable A trion binding energy in the range of 40-100 meV. To further investigate the efficacy of electrostatic control, we compared our measurements with those carried out when the BTO is replaced by a hexagonal boron nitride (hBN) dielectric layer of comparable thickness, confirming BTO's superior gating properties and thus lower power consumption. Additionally, we take advantage of the ferroelectric switching of BTO by fabricating devices where the BTO layer is decoupled from the gate electrode with a SiO$_2$ layer. Choosing to isolate the BTO allows us to induce large remanent behavior of MoS$_2$'s excitonic features, observing hysteretic behavior in the peak energy ratio between A exciton and its trion, as well as hysteretic behavior in the doping-related trion energy shift. This study illustrates the rich physics involved in combining free-standing complex oxide layers with two-dimensional materials.

**INTRODUCTION**

Rapidly after its isolation by mechanical exfoliation in 2005[1] single-layer MoS$_2$ has become an intensely studied material in the realm of two-dimensional (2D) materials for photonics and optoelectronics due to its direct bandgap in the visible range of the spectrum.[2–6] Moreover, single layer MoS$_2$ also has an exciton binding energy unprecedently high for inorganic semiconducting materials that opens the door to study a rich plethora of exciton physics even at room temperature.[7–13] Particularly intriguing is the observation, even at room temperature, of generation of charged excitons, so called trions, that are formed by a neutral exciton and a free electron or hole.[14–18] The generation of trions comes hand-in-hand with a reduction of the population of neutral excitons and can be controlled electrostatically by modifying the density of free carriers in the MoS$_2$ layer.

Up to now, gate dielectrics with a moderate dielectric constant have been used to tune the photoluminescence emission of single-layer MoS$_2$.[7,14,19] In this context, the recent integration of ultrahigh-κ dielectrics based on complex oxides with 2D materials offers a compelling approach for manipulating the optical properties of these systems.[20–23] Although complex oxides like SrTiO$_3$ (STO) or BTO are not van der Waals crystals, they can be isolated to free-standing layers by epitaxially growing them onto a sacrificial layer and proceed to selective etching and peel-off.[24] After the release, those layers can be handled and transferred very similarly to van der Waals materials, providing a powerful new route to fabricate artificial heterostructures combining the rich strongly-correlated physics of complex oxides and the release of lattice-matching need for hetero-stack fabrication.[21,22,25] The interest in these combined heterostructures predates the successful isolation of free-standing oxide layers, with the successful integration of 2D materials and 3D oxide substrate for new functionalities in 2D-based devices.[20,26] Most of these hybrid heterostructures interfaced the 2D material with a ferroelectric oxide film. The spontaneous polarization of the ferroelectric oxide introduces a switchable character to the system, which is especially useful for memristive devices.[27,28] Apart from electronic effects arising from the interaction between the ferroelectric oxide and the 2D material, there have been notable findings regarding light-matter interactions in 2D ferroelectric heterostructures. In exfoliated multilayer $MoS_2$/BTO heterostructures, the optically induced polarization switching phenomenon was initially reported.[29] Subsequent studies have demonstrated that light can serve as an effective tool for advanced programming of various semiconductor-ferroelectric devices.[30] This light-controlled memristive switching capability is being harnessed in the development of efficient optoelectronic synapses for future neuromorphic electronics, with the ability to learn and sense optical information.[31–33] On the other hand, there remains a notable gap in research concerning the influence of the ferroelectric oxides on the optical properties of 2D materials, such as exciton control.[34] Understanding this aspect is of particular interest due to its potential implications for novel memristive and excitonic devices.[12,35–38]

Here, we fabricate hybrid 2D-oxide heterostructures by combining single-layer $MoS_2$ with free-standing $BaTiO_3$ layers, leveraging the ultrahigh dielectric constant of $BaTiO_3$ ($\varepsilon_r \sim 4000$)[25] to control the photoluminescence (PL) emission of the 2D semiconductor through electrostatic gating. Given the large capacitance per unit area and polarization-dependent doping achieved in our BTO/$MoS_2$ heterostructures, we demonstrate a very efficient and strong control of the neutral and charged excitons at room temperature. We also observed an uncommonly large energy difference between the neutral exciton and trion emission peaks (trion binding energy), that is tunable by the applied electrostatic field. Comparing the BTO-based devices to hBN-based ones allows us to define gauge factors for PL emission and trion binding energy tunability. Biasing the monolayer MoS$_2$ through the BTO results in a PL intensity change of 172.8 %V$^{-1}$ and a trion binding energy change of 30.5 meVV$^{-1}$, undoubtingly outperforming the hBN devices with values below 1.4 %V$^{-1}$ of emission change and a non-existing shift of trion binding energy. In addition, by combining BTO with a SiO$_2$ dielectric we take advantage of BTO's ferroelectricity. Employing the combination of dielectrics allows us introduce a remanent behavior to both the peak energy ratio between neutral exciton and trion, as well as their energy difference.

**RESULTS**

Our devices are simple two-material heterostructures of BaTiO$_3$ and MoS$_2$ placed on top of a SiO$_2$/Si substrate with pre-patterned electrodes. Figure 1 shows a sequence of microscopy images of the steps used to assemble the devices under study. We start the fabrication with a SiO$_2$/Si substrate with an oxide thickness of 290nm. The substrate is pre-patterned with buried electrodes, forming a planarized surface with the SiO$_2$ (see cross section of the scheme in Fig. 1d). A 40 nm thick flake of BTO is peeled off from the LSMO/STO substrate, where it has been epitaxially grown, by selective etching of the LSMO that acts as a sacrificial layer (see methods for details).[25] The BTO layer is transferred to a SiO$_2$/Si substrate for further handling. The desired flake is picked up and transferred onto the pre-patterned electrodes using a PDMS/PVC dry transfer technique (Fig. 1a).[39] This method gives full flexibility on flake selection and target position. Subsequently, a single-layer MoS$_2$ flake is mechanically exfoliated from a molybdenite crystal (Molly Hill Mine, Quebec, Canada) onto the surface of a viscoelastic Gel-Film stamp (Gel-Film WF 4 × 6.0 mil by Gel-Pack) and transferred onto the BTO flake surface connected to one of the electrodes (Fig. 1b).[40] This configuration allows us to ground the MoS$_2$ flake while the electrode labelled as "+" is biased to electrostatically gate the MoS$_2$ using the BTO as dielectric layer. An AFM image of the final device (Fig. 1c) reveals the monolayer MoS$_2$ connected to the left electrode labelled as "–" and isolated from the positive electrode by the BTO layer.

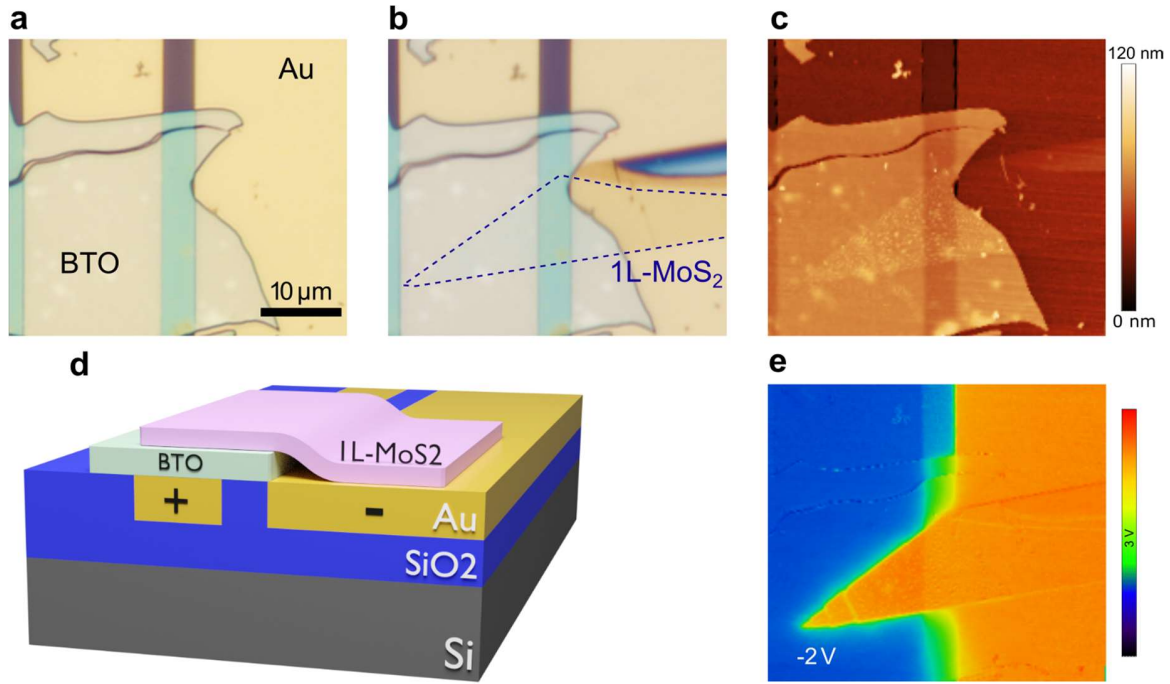

**FIG.1: Device fabrication and configuration.** (a) Substrate with buried electrodes after transfer of BTO. The flake covers the whole area of the bottom electrode (in the image on the left, labelled "+" in (d)). (b) Device after final transfer of single-layer MoS$_2$. The flake is transferred such that it can be connected by the right gold electrode and insulated from the left electrode by BTO. (c) AFM image of the final device structure, revealing the monolayer MoS$_2$ on top of BTO. (d) Schematic of the device, illustrating the buried electrode substrate and the transferred heterostructure on top. The electrodes are labelled for subsequent electrical connections. (e) KPFM in-operando image revealing the surface potential in negatively gated state.

Applying a bias between the two electrodes results in an electric field between the positive electrode and the MoS$_2$ layer. We resolve the built-up potential by mapping the structure using a Kelvin probe force microscope (KPFM) for in-operando measurements. Fig. 1e illustrates the surface potential map for $V_g = -2$ V, obtained by subtracting the $V_g = 0$ V map to eliminate the work function contribution from different regions. The complete set of KPFM images can be found in the supplementary information (Fig. S1). This subtraction technique yields maps of the potential drop across the biased device, revealing a homogeneous surface potential in the MoS$_2$/Au region. It also validates the use of the MoS$_2$ flake as top electrode for polarization switching of the BTO.

In the following, we study the PL emission of our single-layer MoS$_2$ device at room temperature as a function of the gate voltage using a 532 nm laser (see methods for measurement details). Figure 2a shows a comparison of photoluminescence spectra acquired at different biasing conditions and Figure 2b gives a more detailed depiction of the entire sequence of PL spectra taken by color mapping the normalized spectra data. Particularly striking is the tunable emission of excitonic species by the electrostatic field. At negative gate voltages, the PL emission is mostly dominated by the generation of neutral excitons centered around 1.88 eV, as the density of free charge carriers decreases in monolayer MoS$_2$. For positive gate voltages, on the other hand, the emission shows a big contribution due to the generation of trions, negative charged excitons in our case, as expected from an increase of free carrier density induced by electrostatic gating.[14] When no bias is applied the main contribution originates from trions, due to the natural n-type doping of MoS$_2$. This contribution shows up as an extra peak occurring at 1.83 eV. We have performed experiments for different biasing conditions ramping the gate voltage up and down. The intensity and energy of neutral excitons and trions are then determined by fitting the experimental datasets. In order to analyze the PL measurements, we have used pseudo-Voigt functions, which are a linear combination of a Gaussian and a Lorentzian, a common method in analyzing photoluminescence and other spectroscopic techniques.[41–44] The Lorentzian component is linked to the very nature of the emission process, while the Gaussian contribution results from physical factors. Apart from the two peaks related to the A exciton (~1.88 eV) and its trion (~1.83 eV) and a shallow peak with its origin coming from the B exciton (~2.05 eV) we observe another very broad peak with low intensity related to the defect bound excitons of $MoS_2$ (~1.76 eV).[45–47] Thus, we employ four pseudo-Voigt functions to fit the features of our PL measurements with high accuracy (see Fig. S2 for statistics related to the fittings).

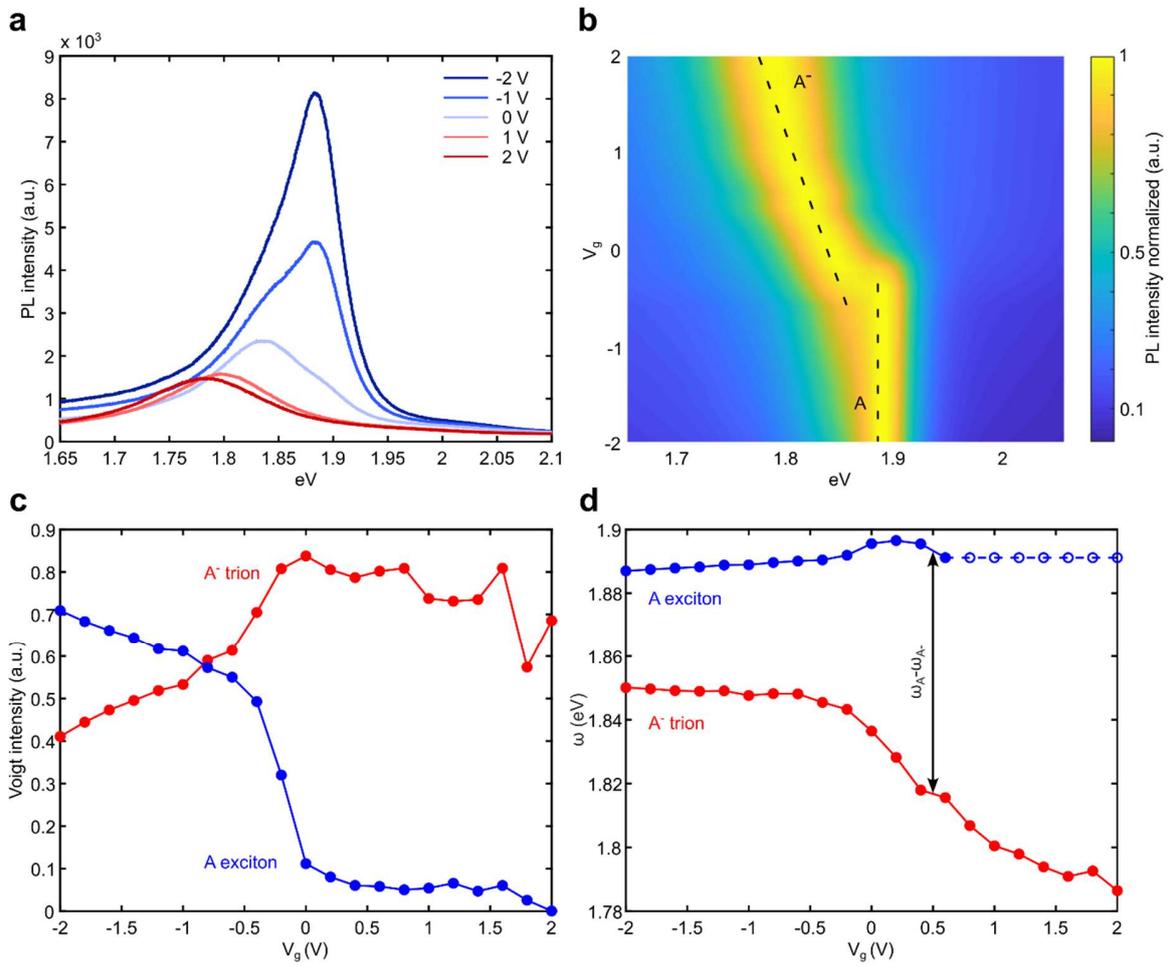

**FIG.2: Tunable emission of MoS2/BTO device at room temperature.** (a) Photoluminescence spectra of $MoS_2$ on top of BTO for different bias voltages. Negative bias increases exciton emission, dominated by the contribution of the neutral exciton. Positive gate voltages increase trion emission through enhancement of free carriers. Furthermore, applying positive gate voltages leads to a progressively larger energy difference between the neutral exciton and trion, which is readily observable in the normalized colormap representation (b). The peak emission energy of the neutral exciton stays roughly the same. However, the trion exhibits a significantly larger shift in peak energy. (c) Intensity changes with gate voltage of the fitted Voigt-functions, illustrating the dominance of the neutral exciton at negative bias and the dominance of the trion at positive bias. (d) Peak energy values of the same fitted functions versus gate voltage, illustrating the clear tuning of trion energy. Data points of the neutral exciton energy at positive gate voltages, where the peak energy is approaching zero, are plotted in empty circles.

The individual fitted functions and overall fits for the data of Fig. 2a, as well as statistics related to the fittings are illustrated in the supporting information (Fig. S2 & S3). The electrostatic influence on the exciton emission of $MoS_2$ is substantial enough to be detectable through optical microscope-based measurements of differential reflectance (Fig. S4). Note, that the Raman spectrum of the $MoS_2$ monolayer does not change with the gate voltage (Fig. S5).

Figure 2c and 2d show a summary of the gate tunable intensities and peak energies for the two exciton species, extracted from the fitted Voigt functions. The figures further highlight the aforementioned behavior between excitons and trions, showing that the two intensities are of equal magnitude at a gate bias of $V_g$ = -0.8 V. Moreover, Figure 2d clearly demonstrates the strong electrostatic dependence of the trion peak energy. As gate voltage increases, so does the energy difference between neutral exciton and its trion. It is important to note that, as the intensity of the A exciton drastically diminishes for positive gate voltages, so does the intensity of the fitted Voigt function, making it difficult to extract the peak energy with high precision. These data points are highlighted as empty circles in Fig. 2d. However, throughout the measurements of a set of devices we confirm the peak energy of the neutral exciton is unaffected by the gate bias, which is supported by other works.[14,19,48]

It is particularly relevant to quantitatively monitor the relative changes between neutral and charged excitons. Figure 3a, for example, plots the intensity ratio between the maximum PL intensity at each gate voltage and the maximum intensity at zero bias, highlighting the strong gate-tunable emission intensity. Noteworthy, the electrostatic control is very efficient: small bias is directly translated to a huge change in photoluminescence intensity due to the ultrahigh dielectric constant of the BTO layer. One can define an emission gauge factor (EGF) as % of intensity change per V of gate voltage to try to quantify the strength of the electrostatic control. This value is relevant as the higher EGF, the lower the necessary power consumption for equal tunability. Figure 3a includes results obtained for the same device configuration with a hexagonal boron-nitride (hBN) of similar thickness (43 nm) instead of BTO to illustrate the strong tuning efficiency of the BTO dielectric against the most common insulating material used with 2D materials. We can extract a specific EGF in the negative gate area of 172.8 %V$^{-1}$ and an overall EGF for the whole voltage range of 20.4 %V$^{-1}$ for the BTO-based device, presenting a high tunability especially in the negative bias regime, with emission intensities more than three times higher compared to zero bias condition. In the case of the hBN-based device it is not possible to identify a reasonable gauge factor as emission intensities do barely vary in the same voltage range, due to the much lower dielectric constant. By increasing the range of applied voltage in the case of the device with hBN dielectric to a range of -10 to 10 V, one can extract an overall EGF of 1.4 %V$^{-1}$, significantly lower than for the BTO-based device (see Fig. S6 for MoS$_2$/hBN device data in extended voltage range). This rather small tunability originates from the fact that in standardly used dielectrics, such as $SiO_2$ or hBN, the trion photoluminescence is largely gate independent, hence the low EGF of around 2 %V$^{-1}$ found in other works based on $SiO_2$.[14,19,49]

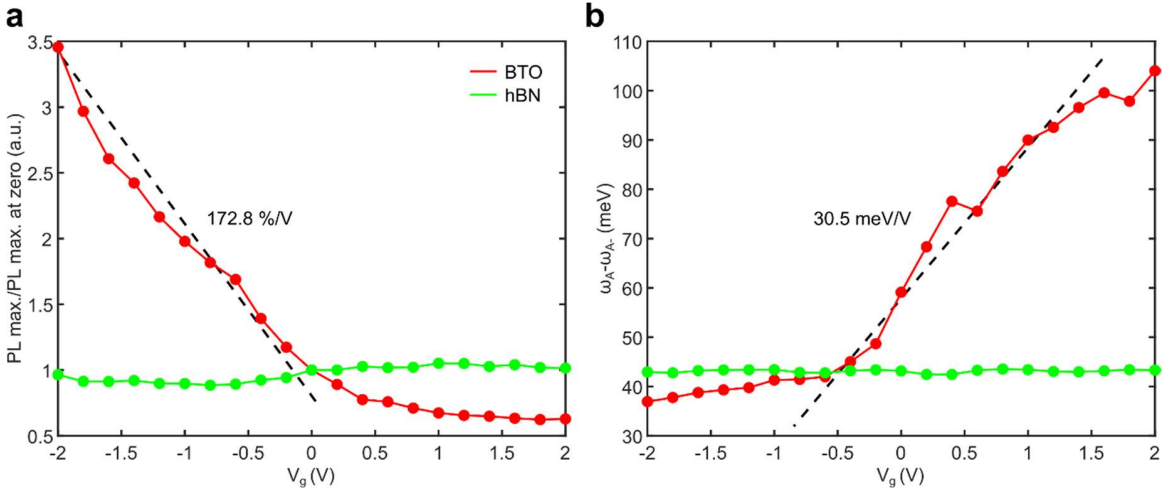

**FIG.3: Exciton emission tuning in Mos$_2$/BTO devices compared to MoS$_2$/hBN devices.** Tuning characteristics for BTO devices are shown in red, while tuning characteristics for hBN devices are shown in green. (a) Gate voltage dependent maximum PL emission intensity related to maximum emission intensity at zero bias for each device. While the emission intensity on the hBN-based device is barely affected by the electrostatic field, the one of the BTO-based device changes significantly, allowing to identify an emission gauge factor of 172.8 %V$^{-1}$. Data is taken directly from the related PL spectra. (b) Gate voltage dependent change of energy difference between neutral A exciton and trion extracted from the fitted Voigt functions, related to the trion binding energy in the monolayer MoS$_2$. Again, while the trion binding energy in the hBN system stays constant, the binding energy for the BTO system is largely influenced by the applied bias. A trion binding energy gauge factor of 30.5 meVV$^{-1}$ shows strong tunability in a small gate voltage window.

The energy difference between the neutral exciton and trion peak is related to the trion binding energy (TBE) by

$$TBE = \omega_A - \omega_{A-},$$

where $\omega_A$ and $\omega_{A-}$ are the peak energies of neutral exciton and trion, respectively. Figure 3b plots this value of the fitted Voigt functions as a function of the applied gate voltage for both the BTO- and hBN-based single-layer MoS$_2$ devices. For the hBN dielectric the trion binding energy is around ~40 meV and almost independent of the bias, in good agreement with pervious works using low- to moderate-κ dielectrics or even suspended $MoS_2$.[7,17,50–52] However, for BTO the binding energy value can be tuned all the way from 40 meV at zero bias to 100 meV at maximum positive bias, almost a factor of 3 larger than for devices based on hBN or $SiO_2$ dielectrics. Again, we try to quantify the extent of tunability in this case for the trion binding energy and define a binding energy gauge factor (BGF) by the change of binding energy in meV per applied V of gate voltage. The BGF in the case of the BTO device is 30.5 meVV$^{-1}$. In the case of the hBN-based device one cannot define such a factor, as there is no change of trion peak position, even for the extended voltage range (see supporting information). Simplified dielectric screening models predict a reduction of trion binding energy for increasing κ[53], which is the opposite to what we observe with BTO. Similar large trion binding energy has been observed using STO as dielectric, however at temperatures lower than 100 K, below STO's transition temperature, and showing binding energies of around 30 meV at room temperature.[54] Although such a drastic change in binding energy has not been shown in any other system, K. F. Mak et al.[14] propose a linear relationship between the Fermi level of $MoS_2$ and the trion binding energy. Results from J. Choi et al.[34] suggest a type I band alignment between $MoS_2$ and BTO, resulting in a large doping of $MoS_2$ and therefore a large free electron concentration in the conduction band. An enhancement of doping and subsequent change in $MoS_2$'s Fermi level is achieved by polarization induced doping, due to the ferroelectric behavior of BTO. Gating with BTO results in large surface charge densities building up to screen polarization charges. We have included measurements of an additional $MoS_2$/BTO heterostructure with similar EGF's in the supporting information (Fig. S7), proving the reproducibility of our system.

One of the key features of BTO is its ferroelectric nature and therefore possible applications in novel memory architectures.[55] Functional characterization of the transferred BTO flakes was performed by Piezoresponse Force microscopy (PFM), to assess the preservation of the ferroelectric character and its switchable nature with applied electric fields. Figure S8 present the hysteresis cycles and the local polarization switching for a BTO sample of similar characteristics as those used for device fabrication. Our findings confirm that the transferred BTO flakes on their own maintain their ferroelectric properties, with observed coercive voltages for domain switching around ±2V. However, while performing the experiments in various devices with the geometry depicted in Figure 1, we did not observe any feature in the PL or hysteretic behavior related to polarization switching and remanence in the BTO. We attribute this to a rapid depolarization of the BTO layer by charge leakage when it is placed between the gold gate electrode and $MoS_2$ top contact. We have observed ferroelectric-induced hysteresis on $MoS_2$ field-effect transistors fabricated on BTO placed onto $SiO_2$/Si substrates in the past.[25] The fast depolarization may stem from free charges in the BTO, due to doping by oxygen vacancies.[56,57] Hence, we designed a new configuration in which we transfer a large flake of BTO, again with a thickness of 40 nm, part of it on top of a gold gate electrode and part on top of $SiO_2$ (290 nm). In this way we can measure two different configurations on the same sample (illustrated in Fig. 4a & Fig. 4d). Firstly, we can repeat the previous measurements by applying the bias between the bottom gold electrode and the $MoS_2$ layer and take the PL measurements on the part of $MoS_2$ that is on top of BTO/Au. And secondly, we can apply the gate voltage between the highly p-doped Silicon substrate and the $MoS_2$, take the PL measurement on the part of $MoS_2$ that is on top of BTO/$SiO_2$, thus isolating the BTO from the bottom gate with SiO$_2$. A more detailed depiction of both device configurations, including a microscope image of the final structure is provided in the supporting information (Fig. S9).

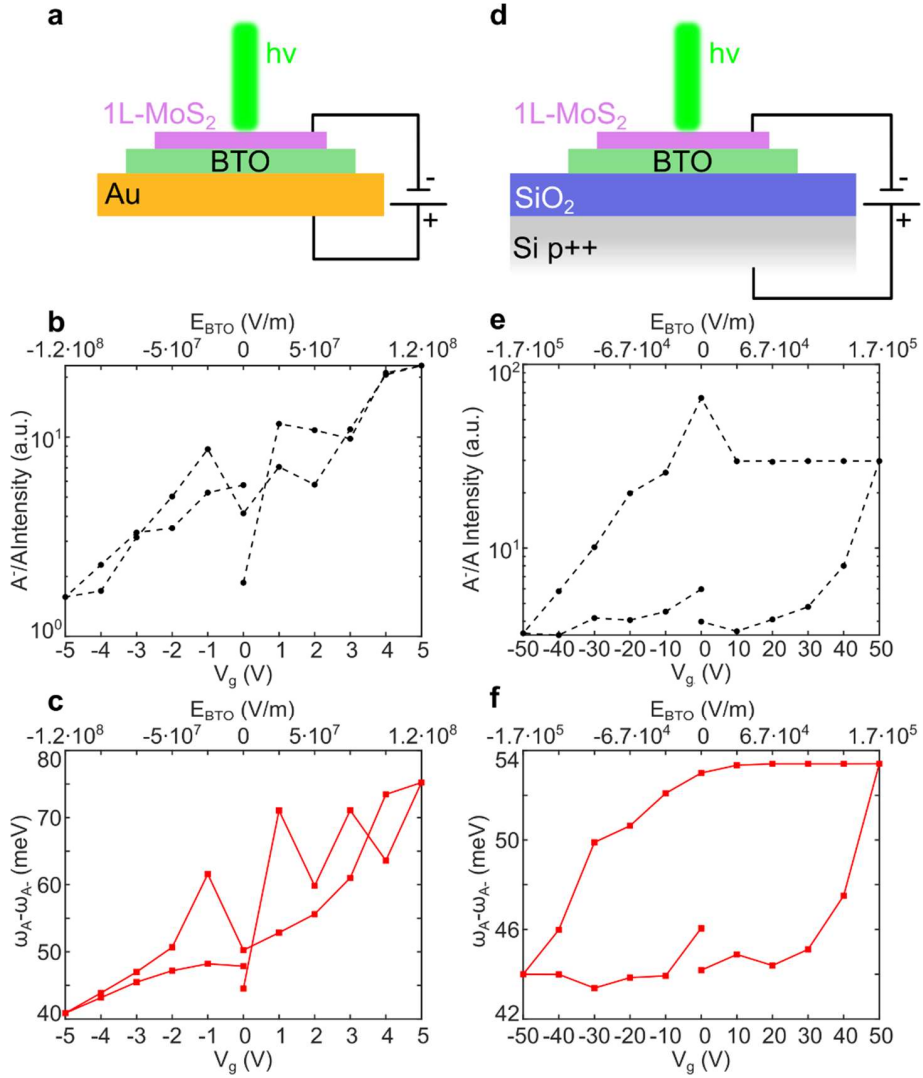

**FIG.4: Dielectric decoupling of BTO for excitonic hysteresis control.** One single device as in Fig. 1d, but two different measurement configurations. In (a)-(c) the measurement configuration coincides with the same configuration used in previous figures 1-3, i.e. gating the MoS$_2$ with only BTO as dielectric layer. However, in (d)-(f) the gate voltage is applied between the highly-doped Si back-gate and the MoS$_2$ monolayer, resulting in a dielectric combination of SiO$_2$ (290 nm) and BTO (40 nm) to exploit the ferroelectric nature of BTO. Schematics of the two configurations are illustrated in (a) &

(d) with a more detailed depiction in the supplementary information. (b) & (e) show the intensity ratio of A trion to neutral exciton in respect to gate bias. (c) & (f) plot the energy difference of their peaks (trion binding energy) for each configuration. (b) & (c) confirm previous results of tunable excitonic features under BTO gating, especially the high tunability of MoS$_2$'s trion binding energy. While no hysteretic behavior is observed in the original configuration, the situation changes when BTO is screened by SiO$_2$. (e) & (f) depict the results for SiO$_2$/BTO gating. In general, tunability decreases in respect to the original situation, however both the ratio of peaks, as well as the trion binding energy now show clear coercivity, depending on the polarization of the BTO layer. The diagrams include a second x-axis with calculated values for the electric field in the BTO layer, to clarify the magnitude of the field necessary to switch polarization using the SiO$_2$ dielectric.

Results for the MoS$_2$/BTO/Au configuration are shown in Fig. 4b-c, this time plotting the ratio between the trion peak and the neutral exciton peak intensity of the fitted functions, as well as the energy difference between the two peaks (TBE of MoS$_2$). Again, this configuration confirms previous claims on large tunability of both features, however doesn't result in remanent behavior. The results for the MoS$_2$/BTO/SiO$_2$/Si configuration on the other hand clearly changes the polarization response of the system and show large coercivity. Fig. 4e-f illustrate the same parameters as for the first configuration, this time showing obvious hysteresis in both situations, the intensity ratio between trion and exciton and curiously as well the trion peak energy and therefore MoS$_2$'s trion binding energy. The introduction of SiO$_2$ to the system decouples the BTO from the gate electrode, prevents charge leakage and therefore clear remanence is observed. The large coercive field in this case can be explained by the series association of the (small) capacitance of the SiO$_2$ oxide and the large capacitance of BTO. Figures 4b,c,e,f include secondary x-axes indicating electric field values within the BTO layer for both configurations, calculated using the simple parallel plate capacitor model (for modelling details see supporting information). Switching in the case of the MoS$_2$/BTO/SiO$_2$/Si configuration is already observed at lower electric field values than those applied to the MoS$_2$/BTO/Au configuration.

**DISCUSSION**

In summary, we present the integration of free-standing layers of BaTiO$_3$ with monolayer MoS$_2$ to electrostatically gate and control the exciton emission of the 2D semiconductor. We show that this configuration leads to highly tunable photoluminescence emission of MoS$_2$ at room temperature, observing large peak emission shifts as well as large changes in the difference of peak emission energies of A exciton and trion, related to the trion binding energy of MoS$_2$. By comparing the BTO-based system with hBN-based devices we can extract gauge factors for maximum emission tuning and for trion binding energy tuning for both systems. This way we showcase the increased tunability of the BTO/MoS$_2$ heterostructures with a very low bias range, and therefore improved power consumption over the hBN counterpart. Additionally, we configure another device structure by implementing BTO with a SiO$_2$ dielectric to make use of BTO's ferroelectric properties. This configuration allows us to create large remanent behavior of MoS$_2$'s excitonic features, resulting in hysteresis of the emission ratio between exciton and trion side by side with hysteresis observation in trion binding energy. Our work demonstrates how the exceptional properties of free-standing complex oxides can lead to an emergence of new features in well-known two-dimensional materials with vast advantages over the commonly used dielectrics for those systems.

**METHODS**

**BTO growth and release**

LSMO (15 nm) and BTO (40nm) films were grown epitaxially onto (001) STO substrates by pure oxygen (3.2 mbar) sputtering technique at high temperature (900 °C). Layer growth was sequential without breaking vacuum. Afterwards the grown heterostructure of LSMO/BTO is adhered to a polydimethylsiloxane film (Gel-Film WF 4 × 6.0 mil by Gel-Pack) and immersed in a diluted solution of 0.5 mL KI (3 mol/L) + 0.5 mL HCl (37%) and 10 mL deionized $H_2O$ at room temperature, dissolving the sacrificial LSMO layer over the course of 3 days and allowing the delamination of the BTO layer without damaging its properties.

**$MoS_2$ exfoliation and transfer**

$MoS_2$ flakes were obtained from natural molybdenite mineral (Molly Hill Mine, Quebec, Canada) by mechanical exfoliation using Nitto tape (Nitto SPV 224) onto a PDMS substrate (Gel-Film WF 4 × 6.0 mil by Gel-Pack). After flake selection and identification with an optical microscope[58] the MoS2 was directly transferred from the PDMS to the final substrate.

**Electrode Deposition**

Prepatterned electrodes were fabricated by mask-less lithography (SmartForce SmartPrint), a 50nm etch of the $SiO_2$ layer (Glass etching cream Armour Etch) and subsequent thermal evaporation of 5 nm Cr (used as adhesion layer) and 45 nm Au onto a $SiO_2$ (290 nm)/Si (p++) substrate, resulting in a planar surface of Au and $SiO_2$.[59]

**Dry PDMS/PVC Transfer of BTO**

The stamp is fabricated by cutting a small squared 5mm x 5mm piece of PDMS (Gel-Film PF 4 × 6.0 mil by Gel-Pack) and placing it onto a glass slide. Next, liquid PDMS is prepared in a ratio of 1:10 (Sylgard 184) and a thin wire is used to consecutively drop three drops of PDMS decreasing in size on top of each other including curing at 100°C on a hotplate for 5min between each drop. As the drop size is decreasing by selectively choosing thinner wires a dome like PDMS pyramid is formed. Finally, a piece of PVC food wrap (Alfapac) is cut and placed on top of the PDMS pyramid using double sided tape. Isopropanol cleaning of the foil is required before starting the transfer. The procedure is adapted from[39] and more details can be found in.[60]

**Electrostatic photoluminescence measurements**

Electrical biasing was realized using a source-meter unit (Keithley 2450) between top and bottom electrodes. Photoluminescence was measured with a confocal Raman microscopy system (MonoVista CRS+ from Spectroscopy & Imaging GmbH) using a 532 nm excitation laser with varying incident power and an 100x short-range objective with a resulting laser spot of around 1 μm in diameter.

**PFM/KPFM Characterization**

A commercial Atomic Force Microscopy (AFM) system, from Nanotec, operating in ambient conditions was employed to perform morphological, piezoresponse (PFM) and surface potential (KPFM) characterization of the samples. Measurements have been acquired in both dynamic (KPFM) and contact (PFM) mode, using the same probe (PtIr-coated commercial tips from Nanosensors (PPP-NCIPt)). For PFM measurements $V_{dc}$ and $V_{ac}$ voltages were applied to the tip while the sample was grounded. The drive frequency, drive amplitude ($V_{ac}$), and trigger force were 52 kHz, 1-3V, and 200-500 nN, respectively. Hysteresis loops were measured in the spectroscopy mode at the selected locations. A sequence of DC voltages was applied to the tip ($V_{dc}$), with the $V_{ac}$ voltage superimposed to excite the electromechanical vibration of the sample. Both, phase and amplitude of the vibration are recorded as a function of $V_{dc}$ voltage. Ferroelectric domain engineering was performed by poling selected areas with $V_{dc}$ tip voltages above the coercive voltages and the $V_{ac}$ modulation off. Subsequent PFM imaging is performed to show the phase contrast due to 180º switching of the polarization. Surface potential maps have been measured using the retrace mode (lift height of 30nm). $V_{ac}$ was applied to the tip, $V_{ac}$ = 1V at a drive frequency of 7 kHz. Image analysis was performed with the Gwyddion free software. A Keithley 2450 source-measure unit was used to apply different bias voltages between the electrodes for in-operando KPFM measurements.

## DATA AVAILIBILTY

The data of this study are available from the corresponding author upon reasonable request.

## COMPETING INTERESTS

The authors declare no conflict of interest.

## ACKNOWLEDGMENTS

We thank the QUDYMA group at ICMM for fruitful discussions. The authors are also thankful to Martin Lee for making us aware of the PVC transfer, which simplified the fabrication process. We acknowledge funding from the EU FLAG-ERA project To2Dox (JTC-2019-009), the Comunidad de Madrid through (MAD2D-CM)-UCM project and the CAIRO-CM project (Y2020/NMT-6661) and the Spanish Ministry of Science and Innovation (Grants PID2020-115566RB-I00, PID2020-118078RB-I00, RTI2018-099054-J-I00, IJC2018-038164-I, TED2021-132267B-I00, TED2021-130196B-C21, PRE2018-084818 and PDC2023-145920-I00).

**AUTHOR INFORMATION**

**Authors and Affiliations**

**2D Foundry research group. Instituto de Ciencia de Materiales de Madrid (ICMM-CSIC), Madrid, Spain**

Thomas Pucher, Sergio Puebla, Estrella Sánchez Viso, Mar Garcia-Hernandez, Carmen Munuera & Andres Castellanos-Gomez

**GFMC, Department Fisica de Materiales, Facultad de Fisica, Universidad Complutense 28040 Madrid, Spain**

Victor Zamora, Victor Rouco, Carlos Leon & Jacobo Santamaria

**Unidad Asociada UCM/CSIC, "Laboratorio de Heteroestructuras con aplicación en spintrónica"**

Carlos Leon, Mar Garcia-Hernandez, Jacobo Santamaria, Carmen Munuera & Andres Castellanos-Gomez

**Corresponding Authors**

Thomas Pucher thomas.pucher@csic.es

Carmen Munuera cmunuera@icmm.csic.es

Andres Castellanos-Gomez andres.castellanos@csic.es

# Supporting Information

# Strong electrostatic control of excitonic features in $MoS_2$ by a free-standing ultrahigh-κ ferroelectric perovskite

*Thomas Pucher[1*], Sergio Puebla[1], Victor Zamora[2], Estrella Sánchez Viso[1], Victor Rouco[2], Carlos Leon[2,3], Mar Garcia-Hernandez[1,3], Jacobo Santamaria[2,3], Carmen Munuera[1,3*] and Andres Castellanos-Gomez[1,3*]*

[1] 2D Foundry research group. Instituto de Ciencia de Materiales de Madrid (ICMM-CSIC), Madrid, 28049, Spain.

[2] GFMC, Department Fisica de Materiales, Facultad de Fisica, Universidad Complutense 28040 Madrid, Spain

[3] Unidad Asociada UCM/CSIC, "Laboratorio de Heteroestructuras con aplicación en spintrónica"

thomas.pucher@csic.es

cmunuera@icmm.csic.es

andres.castellanos@csic.es

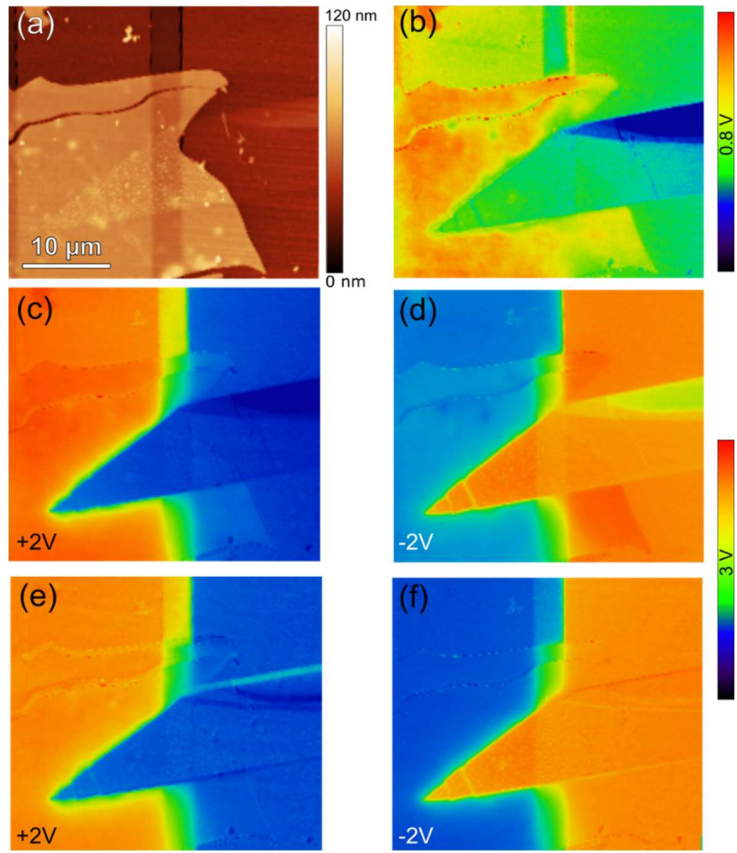

**FIG.S1: Full set of KPFM images for the $MoS_2$/BTO device.** Simultaneous (a) topographic and KPFM images at (b) $V_g = 0$ V, (c) $V_g = +2$ V and (d) $V_g = -2$ V. The bottom row displays subtracted

maps, where the work function contribution (image (b) at $V_g = 0$ V) has been removed, isolating the potential drop contributions for (e) $V_g = +2$ V and (f) $V_g = -2$ V.

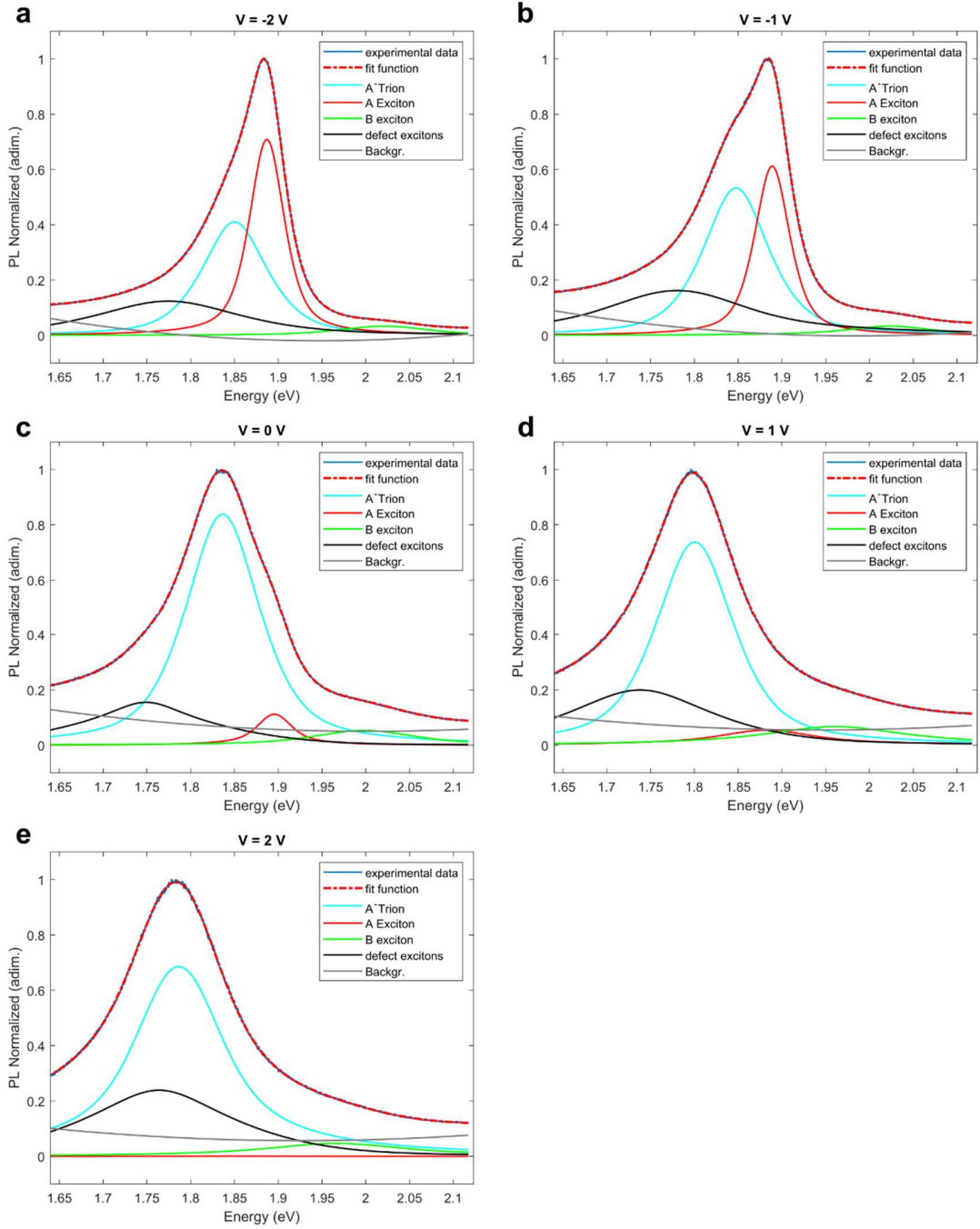

**FIG. S2: PL evolution of MoS$_2$/BTO device by gate voltage including voigt function fittings.** (a)-(e) illustrate the fitted voigt functions of the presented spectra from Figure 2a in the main text. The experimental data is fitted using four pseudo-voigt functions, related to the A exciton, its trion, the B

exciton and lastly the defect excitons, as well as a background subtraction. The dotted red lines represent the overall fit, as the convolution of the individual voigt fit functions, illustrating a well-defined fit for the experimental data.

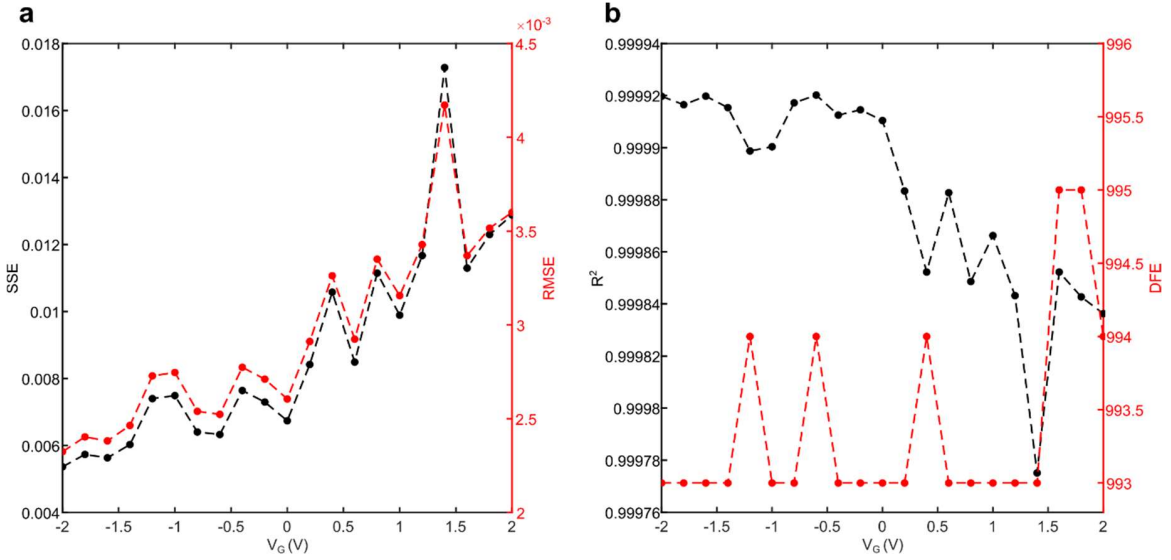

**FIG. S3: Voigt fitting statistics.** (a) Left axis: Sum of Squares Due to Error (SSE), which quantifies the discrepancy between observed values and values predicted by a model, values closer to zero indicates that the model has a smaller random error component. Right axis: Root Mean Square Error (RMSE), which measures the average magnitude of the errors between predicted and observed values, a lower value of RMSE indicates better agreement between the predicted and observed values. (b) Left axis: Coefficient of determination ($R^2$), which represents the proportion of the variance in the dependent variable (the variable being predicted) that is explained by the independent variables (the predictors) in a regression model. Values closer to 1 indicate better model fit. Right axis: Degrees of Freedom for Error (DFE), it quantifies the effective sample size available for estimating the variability of the residuals (errors) in the regression model. Higher values of DFE indicate a greater amount of independent information available for estimating the error variance.

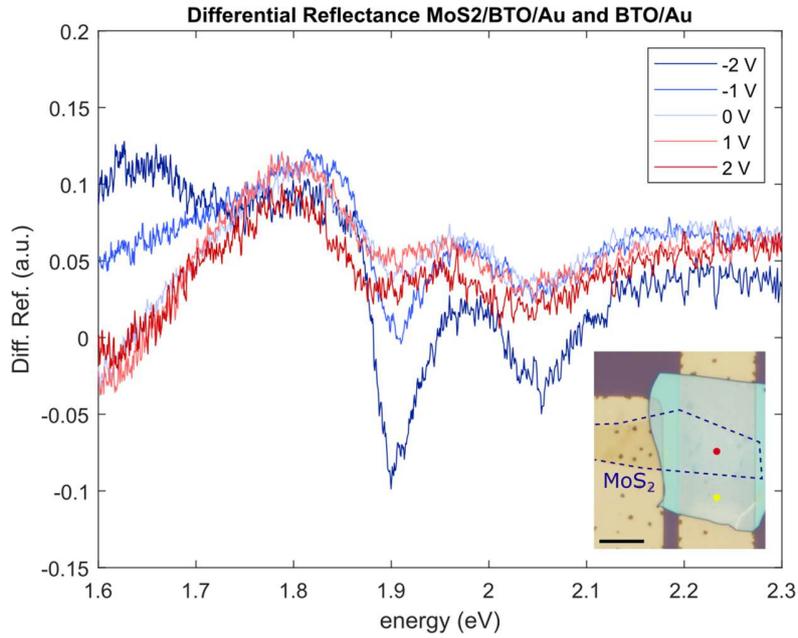

**FIG.S4: Differential Reflectance measurements of MoS$_2$/BTO device.** The inset shows the spots where the reflectance measurements were taken. The red spot corresponds to MoS$_2$/BTO/Au and the yellow spot to BTO/Au configuration. Subtraction of the two gives the differential reflectance depict in the graph for different gate voltages. The scale bar is 10 μm.

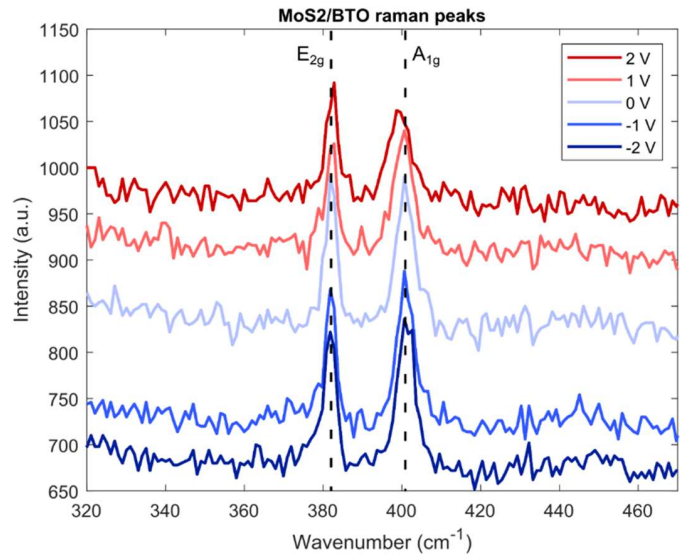

**FIG.S5: Gate voltage dependency of Raman peaks for MoS$_2$/BTO device.** Raman measurements indicate no change in respect to the applied bias. None of the two Raman modes nor the energy difference is affected by the gate voltage. Curves are shown with an offset of 150 a.u. for better visibility.

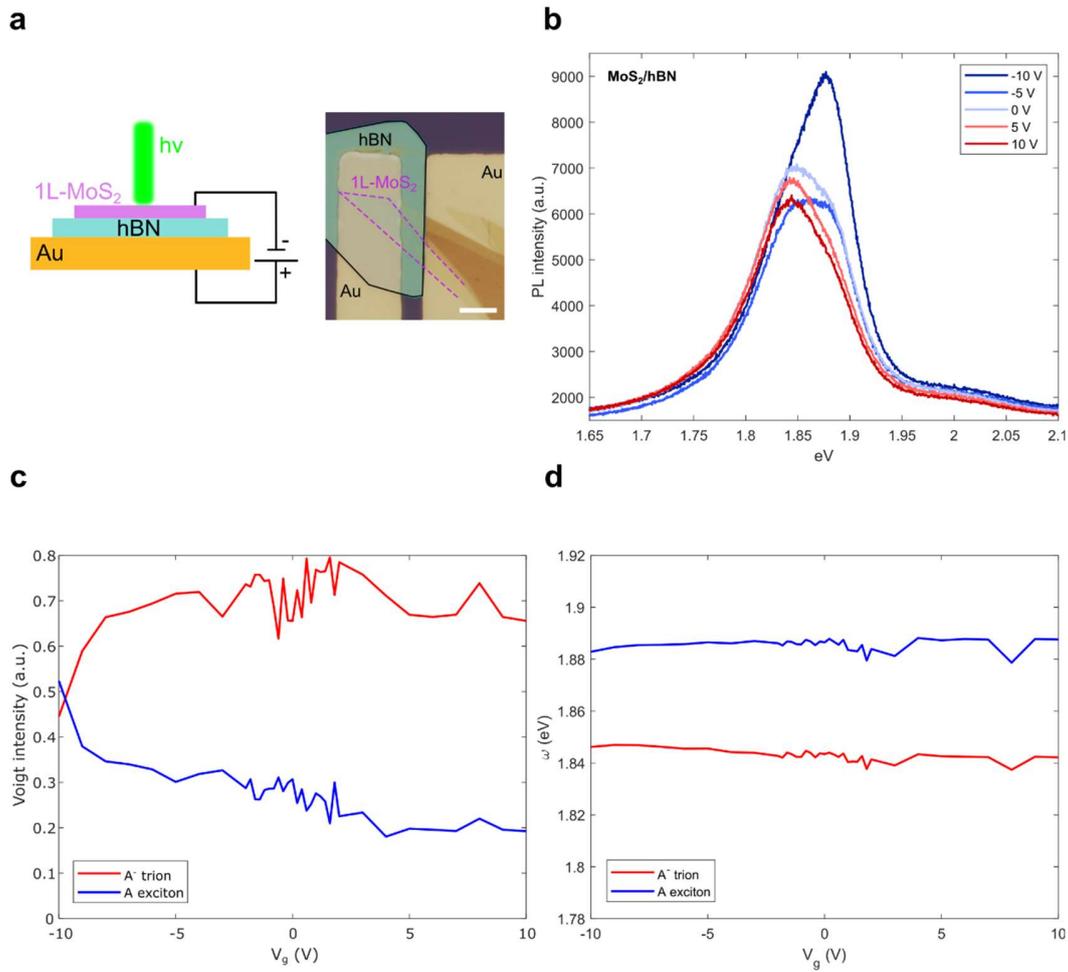

**FIG.S6: Extended data for exciton emission tuning of MoS$_2$/hBN device.** (a) The device configuration is the same as for the MoS$_2$/BTO devices, but instead of BTO hBN, with similar thickness (43 nm) is used as a gate dielectric. The scale bar is 10μm. (b) Electrostatic photoluminescence plots with extended gate voltage range of -10 V to +10 V. Negative gate voltages enhance the emission of the neutral exciton, as expected. Trion emission is not significantly altered by the applied electric field. (c) & (d) Characteristics of the fitted voigt functions for A exciton and trion, comparable to the BTO device data of Fig. 2 in the main text. Even an extended voltage range does not affect the energy difference between the neutral exciton and trion of the MoS$_2$ when gated through hBN.

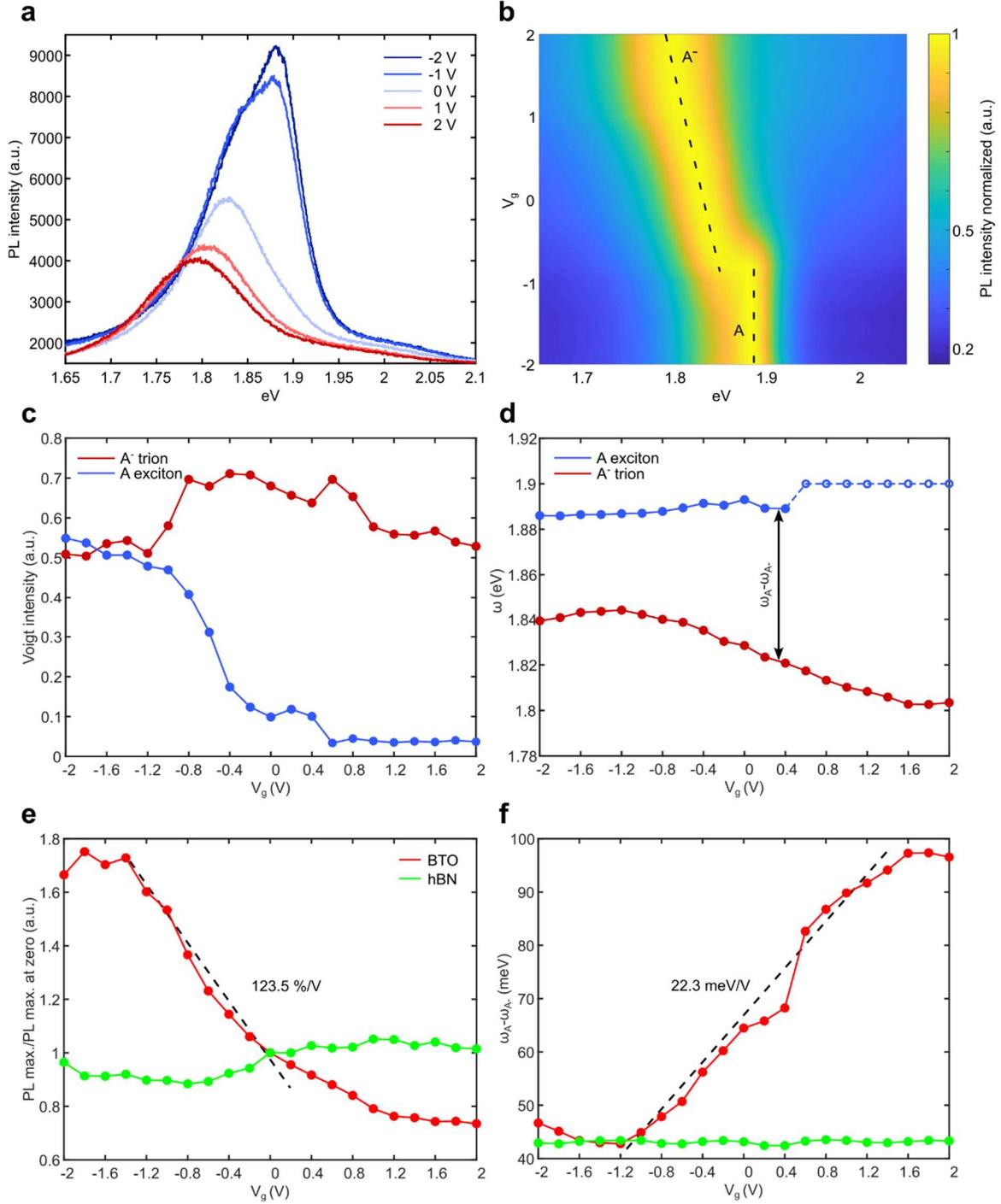

**FIG.S7: Tunable emission of MoS2/BTO sample 2 at room temperature.** (a) Photoluminescence spectra of MoS$_2$ on top of BTO for different bias voltages. Negative bias increases exciton emission, dominated by the contribution of the neutral exciton. Positive gate voltages increase trion emission through enhancement of free carriers. An increases of energy difference between neutral exciton and trion can be observed and gets clear when plotted as colormap in (b). The peak emission energy of the neutral exciton stays roughly the same. However, for the trion we observe a large change in peak

energy. (c) Intensity changes with gate voltage of the fitted Voigt-functions, illustrating the dominance of the neutral exciton at negative bias and the dominance of the trion at positive bias. (d) Peak energy values of the same fitted functions by gate voltage. (e) Gate voltage dependent maximum PL emission intensity related to maximum emission intensity at zero bias for each device. While the emission intensity on the hBN-based device is barely affected by the electrostatic field, the one of the BTO-based device changes significantly, allowing to identify a specific EGF of 123.5 %V$^{-1}$ and an overall EGF of 16.9 %V$^{-1}$. Data is taken directly from the related PL spectra. (f) Gate voltage dependent change of energy difference between neutral A exciton and trion extracted from the fitted voigt functions, related to the trion binding energy in the monolayer MoS$_2$. Again, while the trion binding energy in the hBN system stays constant, the binding energy for the BTO system is largely influenced by the applied bias. A trion binding energy gauge factor of 22.3 meVV$^{-1}$ shows strong tunability in a small gate voltage window.

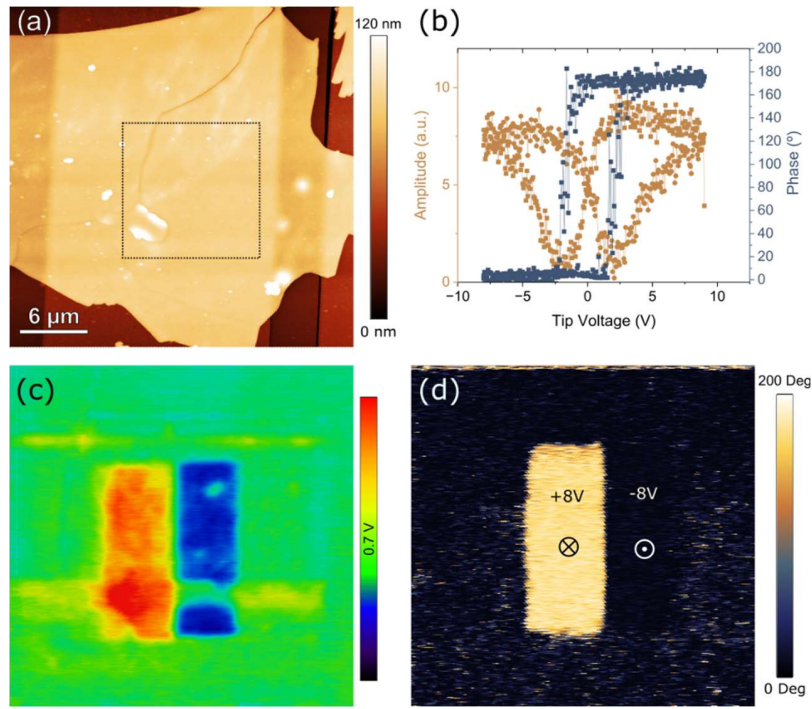

**FIG.S8: BTO PFM measurements.** (a) Topographic image of a 40 nm thick BTO flake, transferred on top of a buried Au electrode. (b) Amplitude (orange) and phase (blue) local PFM hysteresis loops acquired on the transferred BTO flake. (c) KPFM and (d) PFM phase images after ferroelectric domain engineering by pooling the square area marked in (a) with tip voltages of +8V and -8V. PFM imaging shows the phase contrast due to 180º switching of the polarization. The KPFM image shows a decrease in the surface potential for the upward polarization (in agreement with accumulation of negative charges at the BTO surface), in comparison with the value measured for the downward polarization.

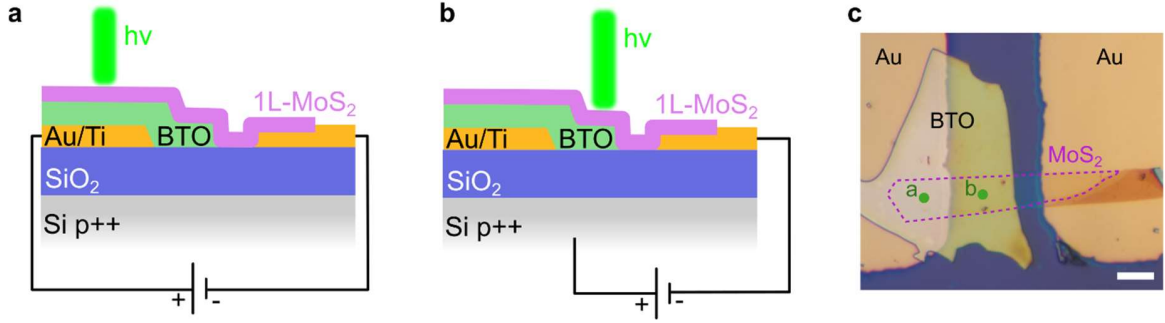

**FIG.S9: Detailed device configuration of MoS$_2$/BTO/SiO$_2$ devices.** (a) & (b) illustrate the device configurations of Fig. 4 of the main text in more detail. The BTO flake was transferred in such a way that half of the BTO is covering the left gold electrode and the other half is situated on the SiO$_2$ (290 nm) area without gold underneath. (c) Optical microscope image of the device outlining the monolayer MoS$_2$ and the two different laser spots for the measurement configurations of (a) and (b). In configuration (a) the device is gated between the left and the right gold electrode, resulting in gating the MoS$_2$ only through the BTO. In configuration (b) the bias is applied between the bottom highly p-doped Si and the right gold electrode, resulting in gating the MoS$_2$ through BTO/SiO$_2$. The scale bar is 10 μm.

**Electric field calculations for both configurations**

For the configuration of Fig. S9a the determination of electric field is as simple as

$$E_{BTO} = \frac{V}{t_{BTO}},$$

where V is the applied bias and $t_{BTO}$ is the thickness of the BTO layer (40 nm).

To extract the electric field in the BTO layer for the second configuration of Fig. S9b, the overall capacitance of the two series capacitances is dominated by the SiO$_2$ layer, as BTO's dielectric constant is orders of magnitude larger than that of SiO$_2$ ($\varepsilon_{r,BTO}$ ~ 4000, $\varepsilon_{r,SiO2}$ = 3.9). The electric field for this configuration can therefore be calculated by

$$E_{BTO} = \frac{C'_{SiO2} \cdot V}{\varepsilon_0 \cdot \varepsilon_{r,BTO}},$$

Where C'$_{SiO2}$ is the capacitance per area for the SiO$_2$ layer with a thickness of 290 nm, V is the applied bias and ε$_0$ the vacuum permittivity ($8.85 \cdot 10^{-12}\ Fm^{-1}$).